# A possible explanation of the phase diagram of cuprate superconductors

Tian De Cao*, Biao Son, Chuan Qi Li, Tie Bang Wang, Sheng Li Guo

*Department of physics, Nanjing University of Information Science & Technology, Nanjing 210044, China*

A d-p pairing curve that is consistent with the pseudogap curve observed in experiments is found on a d-p model on phonon mechanism. On the discovery we suggest that there are two pseudogaps associated with the nearly localized d-p pairs and nearly free p-p pairs. The p-p pairs look like bosons and are responsible for superconductivities.

PACS number(s): 74.20.-z

Keywords: d-p pairing; p-p pairing; pseudogap; diagram.

## Ⅰ. INTRODUCTION

The last decade witnessed significant progress in experiments of high temperature superconductors, especially in the observation of pseudogap. The so-called pseudogap state [1-4] is formed in the normal state that resembles the superconducting gap, in both magnitude and symmetry, leading to a view that the origin of the pseudogap may be intimately related to superconducting pairing at $T > T_c$. An alternate point of view asserts that the pseudogap represents a state that competes with superconductivity. Hence the relation between the pseudogap and superconductivity has been a riddle.

To understand the pseudogaps, we must know the pairing temperature. Although someone suggested that the pairing temperature might be higher than the superconducting temperature, almost all authors have taken the pairing temperature as the superconducting one in literatures, so do I in past paper [5]. In that paper the high $T_c$ associated with d-p pairing was suggested. Nowadays, we find that the pairing temperature may be head and shoulders above the superconducting temperature and that the d-p pairs are usually localized from light- to under- or even optimal-doped regions, hence the so-called high-$T_c$ related to superconducting mechanisms argued by some authors in literatures may have to be questioned. In this paper, we will focus on the origin of the pseudogap.

## Ⅱ. LOCALIZED D-P PAIRS ASSOCIATED WITH PSEUDOGAP

With the hole notation, it is suggested that the d-p model upon low doping is the full description of these $3d_{x^2-y^2}$ and $2p_\sigma$ orbitals in the copper oxides where the Hamiltonian takes the forms

$$H_{dp} = H_{dpt} + H_{dpU} + H_{dpV}, \qquad (1)$$

$$H_{dpt} = \varepsilon_d \sum_{i\sigma} d^+_{i\sigma} d_{i\sigma} + \varepsilon_p \sum_{l\sigma} p^+_{l\sigma} p_{l\sigma} - \sum_{<il>\sigma} t_{il}(d^+_{i\sigma} p_{l\sigma} + h.c.),$$

$$H_{dpU} = U_d \sum_i n_{i\uparrow} n_{i\downarrow},$$

$$H_{dpV} = V_{dp} \sum_{<il>} n_i n_l.$$

At present, the model is so complex that some stepwise handles must be taken. The overlap terms $t_{il}$ may lead to the hop $d_i \to d_{i'}$, so the features of d-holes may be included in the Hubbard model

$$H_d = \sum_{<i,i'>\sigma} \varepsilon_d(i,i') d_{i\sigma}^+ d_{i'\sigma} + U_d \sum_i n_{i\uparrow} n_{i\downarrow}, \tag{2}$$

and the d-holes are localized when the model is half-filled for large $U_d$. This makes us understand that the d-holes are also localized on model (1) upon low doping if $U_d \gg \varepsilon_p - \varepsilon_d \gg t_{il}$. Therefore, when we consider the dynamics of p-holes upon low doping, the model (1) might be simply deduced to the Hamiltonian

$$H_{low-doping} = \sum_{\vec{k},\sigma} \xi_d d_{\vec{k}\sigma}^+ d_{\vec{k}\sigma} + \sum_{\vec{k},\sigma} \xi_{p\vec{k}} p_{\vec{k}\sigma}^+ p_{\vec{k}\sigma} + \sum_{\substack{\vec{k},\vec{k}',\vec{q} \\ \sigma,\sigma'}} V_{\vec{q}} d_{\vec{k}+\vec{q}\sigma}^+ d_{\vec{k}\sigma} p_{\vec{k}'-\vec{q}\sigma'}^+ p_{\vec{k}'\sigma'}, \tag{3}$$

with the constraint $n_{id} = 1$.

In (3) we proposed that the d-holes form an antiferromagnetic ground in $CuO_2$ plane. The model (3) looks like the exciton model of semi-conductors except the constraint $n_{id} = 1$. In semi conductors the exciton states are formed in their band-gap. Here we will find that the model (3) lead the d-p pairing states to lying within the charge-transfer gap $\Delta_g = \varepsilon_p - \varepsilon_d$. Because the d-p pairing gap (as discussed below) $\Delta_{dp} \ll \Delta_g$, the d-p pairing demands having in-gap states. In paper [5] we found the in-gap states can appear in the model similar to (3). However, in this paper the roles of in-gap states will be replaced by the changes of the gap: $\Delta_g \to \Delta_g' < \Delta_g$ when the effect of strong correlation is partly considered.

Considering the lightly doping limit, we can neglect the dependences of $\xi_{p\vec{k}}$ $(=\varepsilon_{p\vec{k}} - \mu)$ on wave vectors $\vec{k}$, and then we will extend our results to other doping region. As the calculating process in paper [5], we find the p-p pairing function $F_p^+$ must be zero if the d-p pairing function $F_{dp}^+ = 0$. Therefore, we get:

**Conclusion 1:** the d-p pairing temperature (defined as $T_1^*$) must be larger than the p-p pairing temperature (defined as $T_2^*$) upon low doping.

Nearby the temperature $T \sim T_1^*$ we get the d-p pairing-gap equations

$$\Delta_{dp}(\vec{k},\sigma,\tau=0) = -\frac{n_F(-\xi_d - V_0 x/2) - n_F(\xi_p + V_0/2)}{\xi_p + \xi_d + V_0 x/2 + V_0/2} \sum_{\vec{q}} V_{\vec{q}} \Delta^+_{dp}(-\vec{k}-\vec{q},\sigma,\tau=0), \quad (4)$$

where the d-p gap function is defined as $\Delta^+_{dp}(\vec{k},\sigma) = \sum_{\vec{q}} V_{\vec{q}} F^+_{dp}(\vec{k}-\vec{q},\sigma,\tau=0)$, and we have taken $n_{di\uparrow} = n_{di\downarrow} = 1/2$ that is consistent with the antiferromagnetic ground formed by d-holes. The d-p pairs intend to destroy the antiferromagnetic order, so there is only short length order with increased doping, for example, as the case in underdoped region. We also take $n_{pl}$ (the holes per oxygen-site) as $x$ (the doping concentration of holes) in (4). In low doping region for model (3), the chemical potential should be located at the position defined by $-\xi_d - V_0 x_0/2 = \xi_p + V_0/2 = 0$. Letting $x = x_0 + \Delta x$, we have

$$\Delta^+_{dp}(-\vec{k},\bar{\sigma},\tau=0) = -\frac{n_F(-V_0\Delta x/2) - n_F(0)}{V_0\Delta x/2} \times \sum_{\vec{q}} V_{\vec{q}} \Delta^+_{dp}(-\vec{k}-\vec{q},\sigma,\tau=0). \quad (5)$$

For the d-p pairing gap function, for example, if the $V_{\vec{q}}$ is relatively attractive for the small $\vec{q}$, and repulsive for the large $\vec{q}$, the predominantly $d$-wave d-p pairing can be obtained in our results. On the equation (5), we evaluate that the $T^*_1 - x$ curve is approximately decided by

$$(x-x_0)^2 = \frac{4}{V_0^2} T_1^{*2} + \frac{32}{V_0^3} C T_1^{*3}, \quad (6)$$

where $C$ is a positive constant number. If $V_0 > 0$, that is, when affective d-p interaction is repulsive, we evaluate $T^*_1 \sim 0K$ by Expression (5), no d-p pairing. Hence we get:

**Conclusion 2:** Pure repulsive interaction between carriers cannot lead to being paired.

This conclusion may lead to some debating since many people (starting with a famous Physical Review Letters by Luttinger and Kohn published in 1965[6]) have proposed pairing mechanisms based on ordinary repulsive interactions. We suppose $V_{\vec{q}}$ is relatively attractive for the small $\vec{q}$, and repulsive for the large $\vec{q}$ following Bulut and Scalapino's suggestion [7] under the phonon mechanism. As $T^*_1 \geq T^*_2$, we obtain the qualitative results in Fig.1 with low doping. In this figure, we take $x_0 \sim 0$.

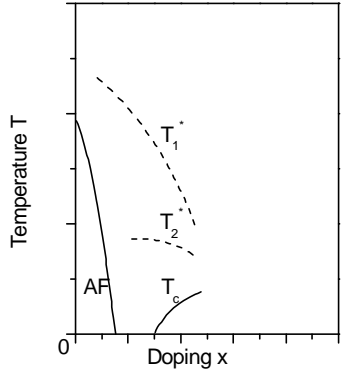

FIG.1. Pseudogap curve and superconducting curve of p-type superconductors are obtained with the d-p interaction upon light doping. $T_c$ is the superconducting temperature associated with p-p pairing.

Some parts of the phase diagram of p-type superconductors are shown in Fig.1 with d-p interaction. In this figure, we suppose that the d-p pairs are not carriers, but the p-p pairs are charge carriers. However, with increased doping, the p-p interaction must be more and more important, the model (3) will be not affective for cuprates. In this case, the mobile p-p pairs look like bosons, this means that the transition temperature must be the Bose-Einstein condensation temperature, thus $T_c$ can be approximately related to $x_{pair}$ (the number density of p-p pairs) as the forms

$$T_c = \frac{\pi}{(5.224)^{2/3}} \frac{\hbar^2}{m_h k_B} x_{pair}^{2/3}. \tag{7}$$

It is found $x_{pair} \neq 0$ demands $T \leq T_2^*$. Because the p-p pairing is the vanguard of superconductivity, $T_c = T_2^*$ demands enough large $x_{pair}$, hence $T_c \leq T_2^*$ in all doped section. In extremely lightly doped region, the p-p pairs are little and localized, $T_c = 0$. In lightly doped region, the number density of p-p pairs is also small, we also get $T_c \ll T_2^*$ and $T_c \to 0$, hence we have Fig.1.

## III. A NEW PHASE DIAGRAM

On the basis of the evaluation above and compared with experiments, we can predict and explain the phase diagram shown in Fig.2. The phase diagrams should vary with different cuprate materials.

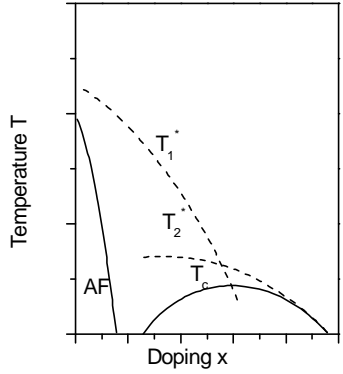

FIG.2. Schematic phase diagram of p-type superconductors is suggested with the phonon mechanism.

It is necessary to note that the p-p pairing curve may be not the bell-shaped one. We believe that the van Hove singularity of electron states is near the Fermi surface in optimally doped region but is far and far from the Fermi surface in increased doping. Moreover, the hole-hole interaction must intend to reduce the electron-phonon interaction, $T_2^* \to 0$ with increased doping, hence superconducting curve is the bell-shaped one. These lead us to get the results:

**Conclusion 3:** The plot of $T_c$ vs carrier concentration shows the characteristic bell-shaped dependence in superconducting region.

However, what the pseudogap curves were suggested by experiments is shown in Fig.3, where the p-p pairing curves have not been presented in any literature.

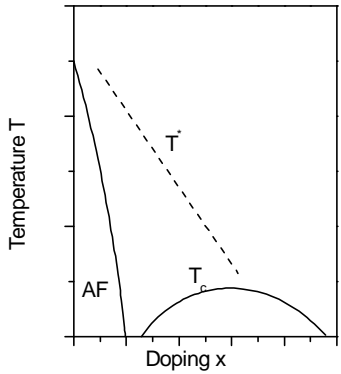

FIG3. Schematic phase diagram of p-type superconductors was suggested with experiments.

We assert that the phase diagram of Fig.3 should be corrected into the one of Fig.2. Someone may question our discussion because they question the relations between phonons and properties of cuprates.

Some properties are explained with the van Hove singularity by authors, such as the explanation of high-$T_c$ [8,9]. However, some authors argued against the van Hove singularity scenario for cuprate superconductivity, for example, A. Kaminski and his coauthors compared the results in LaSrCuo and in BiSrCuO [10], this may be because these scenarios did not well handle the roles of strongly correlation. The van Hove singularity has the evidences of experiments [11]. We find the low frequency phonons exchanged between holes dominate the physics of cuprate superconductors owing to the van Hove singularity of electron states in optimally doped region, and most properties of cuprates can be understood on the phonon mechanism (for example, please see arXiv.org cond-mat\0609339).

**REFERENCES**

[1] W. W. Warren, Jr., R. E. Walstedt, G. F. Brennert, et al, Phys. Rev. Lett. 62(1989)1193-1196.
[2] Y. Yoshinari, H. Yasuoka, Y. Ueda, K. Koga, and K. Kosuge, J. Phys. Soc. Jpn. 59(1990) 3698.
[3] D. N. Basov, T. Timusk, B. Dabrowski, and J. D. Jorgensen, Phys. Rev. B 50(1994) 3511-3514.
[4] Ienari Iguchi, Tetsuji Yamaguchi, and Akira Sugimoto, Nature 412(2001)420.
[5] T. D. Cao, S. L. Guo, C. Q. Li, G. S. Cheng, Physica C402(2004)388.
[6] W. Kohn, J. M. Luttinge, Phys. Rev. Lett. 15(1965)524.
[7] N. Bulut, D. J. Scalapino, Phys. Rev. B54 (1996)14971.
[8] J. Friedel, J. Physique 48(1987)1787.
[9] J. Labbe, and J. Bok, Europhys Lett3(1987)1225.
[10] A. Kaminski, S. Rosenkranz, H. M. Fretwell, et al, Phys. Rev. B 73, e174511(2006).
[11] K. Gofron, J. C. Campuzano, A. A. Abrikosov, et al, Phys. Rev. Lett. 73(1994)3302.